# Recrystallization Characteristics of Catalytic Alloy and Graphite in Diamond Synthesis

Sang Jun Cha, Myong Chol Pak*, Kwang-Il Kim, and Su Gon Kim
Department of Physics, **Kim Il Sung** University, Ryongnam Dong, Taesong District
Pyongyang, Democratic People's Republic of Korea

We first consider the recrystallization characteristics of catalysis alloy and graphite in the process of diamond synthesis under the condition of super high pressure and high temperature in catalysis method. In the process of diamond synthesis catalysis metal is plastically deformed by increase of pressure and then recrystallized as increasing the temperature. As catalysis metal is recrystallized, the shape of graphite particle is in spherical shape in the region contacting with the catalyst but in any shape in the opposite region.

In addition, we calculate the electron charge density distribution and cohesive energies of cementite structure using the first principle method to investigate the reciprocal interaction between transient metal elements and carbon atoms in high-temperature catalyst synthesis. After determination of lattice constant parameters, we obtain the cohesive energy by subtracting the total energy of the crystal from the summation of total energies of atoms composing the crystal and dividing it by the number of atoms. Therefore, the effect of the catalyst on the diamond synthesis is to be analyzed theoretically.

## 1. Introduction

Most of the industrial diamond crystals nowadays are synthetic [1, 2]. Diamond can be synthesized by various methods [3, 4]. The equilibrium pressure and temperature for the transformation from graphite to diamond are very high [5] . Because of the successful use of metallic solutions the transformation pressure/temperature is reduced. Strong and Hanneman had proposed a phase diagram for carbon–nickel system [6]. They indicated that the four-phase (solid–liquid–diamond–graphite) invariant point for nickel–carbon system is 1665 K and 5.25 GPa. The diamond can thus be prepared above this point by using metallic nickel–carbon solutions. In addition to nickel, boron is added to synthesize thermally stable diamond crystals [7], and the use of additive h-BN as a catalyst is also proceeding to synthesize strip-shaped diamond crystals peculiar to Fe-Ni-C system [8]. It also discusses the role of sulfur, an additive when synthesizing industrial diamond with an alloy powder of $Ni_{70}Mn_{25}Co_5$ as a catalyst. The effects of zinc additives on nucleation and growth of diamond have also been studied [11]. They have found that only moderate addition of zinc to conventional catalysts can accelerate the growth of diamond, and rather, excessive addition inhibits nucleation and growth. In addition, it has been studied that even when using Ni-based catalysts, crystallization of crystals with very high nitrogen concentration can be made by using temperature gradient method [12]. Also, other metals such as Fe, Co, Pt, Pa, Rh, Ir, Os, Ta, Mn and Cr have also been used for the synthesis of diamond crystals [13,14,15]. There is also an active research project on the effect of Cu-based alloys on the synthesis of single crystal steels [16]. They found that adding copper to the catalyst could not lower the reaction temperature required for diamond growth. In recent years, studies have also been conducted to synthesize diamondite using the rare earth elements as catalyst [2].

The quality of diamond synthesized by catalysis method depends on type and characteristics of a catalysis metal [17]. The catalytic transition metal elements that are now the most widely used in diamond synthesis are Fe, Ni, Co, Mn and Cr. Catalysis metal element Fe, Ni, Co, Mn form metastable carbides with a structure similar to that of $Fe_3C$ where each carbon is surrounded by six iron atoms in the process of diamond synthesis[18]. In the Ref. [19], it proposed a combinative mechanism of HP-HT(high pressure and high temperature) catalytic synthesis of diamond based on X-ray data obtained from synchrotron radiation and other experimental results. In this mechanism, thay suggested that metal atoms of the catalyst-solvent diffuse between the graphite layers to form weakly bonded graphite intercalation compounds(GICs). However, previous researchers have not mentioned the recrystallization characteristics of the catalyst alloys and the graphite recrystallization properties in contact with them during the HP-HT catalytic process.

Therefore, we are going to study the recrystallization characteristics of catalytic metal and graphite in the synthesis process of diamond with catalysis under HP-HT. Additionlly, we are going to consider the typical transition metal carbides of cementite structure ($M_3C$: M for metal, space group of Pnma) to investigate the interaction characteristics between transition metal elements and carbon atom. Based on the pseudopotential plane wave (PP-PW) method and the projector augmented wave (PAW) method within density functional theory (DFT)[20], the crystal structure of carbides containing Mn, Fe, Co, and Ni, which are transient metal elements, is simulated, compared with experimental data, and the cohesive energy is calculated to reveal the effect of the catalyst on chemical bonding.

This paper is organized as follows. In Section 2, we give an experimental method to clarify the recrystallization characteristics of catalytic metals and graphite when the diamond is synthesized by catalytic method under HP-HT and analyze the results. In Section 3, the first principle method is used to determine the crystal lattice structure and cohesive energy of 3d transition metal carbides and to investigate the interaction between transition metal elements and carbon atoms in the diamond synthesis.

## 2. Experiment Method and Result Analysis

The super high press used in the experiment was a wound-type 2,000 ton super high press.
The device for producing super high pressure was a gouge-anvil type, high-purity graphite was used and the catalysis was $Ni_{40}Mn_{30}Fe_{30}$, $Ni_{70}Mn_{25}Co_5$.

The pressure inside a super high pressure chamber was measured by using PbSe(4.3GPa), CdTe(3.5GPa), the temperature by a Kromel-Alumel thermoelectric thermometer, and the error generated by pressure-heat electromotive force effect was corrected. Pressure transmitting materials is pyrophyllite.

In the experiment we elucidated a metallic behavior of catalysis plane, which was processed as diamond synthesis under the super high pressure and high temperature, and the recrystallization behavior of graphite plane, which was contacted with the catalysis plane, by metallic microscopic and scanning electron microscopic observation.

### 2.1. Recrystallization of catalysis alloy under HP-HT

The metal structure of the catalyst affects the synthesis of diamond. The size of crystalline particle, number of crystalline system, stress status of metallic organization, and bonding concentration etc., effect on the melting point of the alloy as well as nucleus generation and growth of diamond.

Because the plane catalysis was processed through hot rolling and cold rolling, the organization status of metal was very severely changed.

Among catalysis planes processed through diamond synthesis we selected the catalysis plane below of above the reagent chamber, where the temperature of synthesis chamber is relatively low, as a sample under consideration. In the centre of catalysis plane selected as a coupon diamond was synthesized, while in the edge diamond was not synthesized because the temperature is relatively low there.

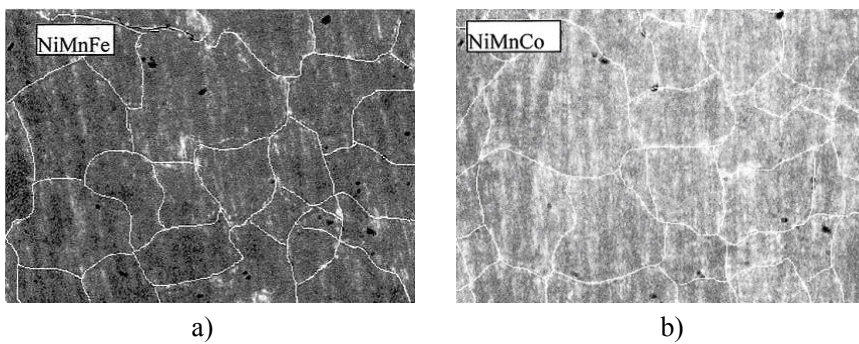

a)                                      b)

**Figure 1. Metallic microscopic photo of catalytic alloy after synthesis**

Mirror the area where the diamond was not formed, and then corrode the specimen with nitro hydrochloric acid supersaturated with copper chloride. And then we observed organization status, modification status and particle size by using metallic microscope "NEOPHOT-32". At this time we estimated the average size of particle by pointing as the same ratio with $0.01\times10^{-3}$m standard ruler. The magnification ratio is 400 when doing metallic microscopic observation. (Fig. 1) As shown in the microscope picture the organization of catalytic metal was recrystallized as average size 80 μm of particle, the organization shape is close to polygon and the crystalline interface was very clear. Such phenomena were observed in both catalytic metals. Therefore, we can conclude that the catalytic metal is firstly calcined, broken as small sized particles by increase of pressure in diamond synthesis, and then recrystallized as increasing the temperature. Thus, it is clear that the control of temperature-pressure condition when diamond synthesis by catalytic method is very important problem.

## 2.2. Recrystallization property of graphite under HP-HT

The recrystallization process occurs on the graphite plate contacting with a catalysis plate. The catalytic alloy facilitated the recrystallization process of graphite. This is that's why strain energy reserved in the catalytic metal plate is released during the recrystallization process and this energy just facilitates the recrystallization process of graphite.

The recrystallization of graphite can be realized when the condition of high pressure and high temperature over 2000 ℃ is retained for several hours. The synthesis regime of diamond, however, is several hundreds of Kelvin lower than this temperature. Graphite is also strained very much at high temperature, where the strain energy is much lower than that in the catalytic metal.

When applying pressure, cluster of equilibrium layer of carbon hexagon net plane is quite stained and graphite particles of 26 nm size are recrystallized. The recrystallization of graphite increases by strain energy reserved within the graphite crystal and furthermore it is facilitated by energy released during the recrystallization of catalytic metal. This recrystallization of both graphite and catalytic metal above mentioned occurs before melting of catalytic metal.

Through the scanning electronic microscopy (SEM) photos of cutting section of graphite, a plane contacting with catalysis and the opposite plane under the high temperature of 1400~1600 ℃ and high pressure of 4.5 ~ 5 GPa, and retaining time of 1 minute, the behavior of graphite under high pressure and high temperature can be understood. In Figure 2 the SEM photos of graphite processed under high pressure and high temperature are shown. The magnification was 800 times.. The scanning electronic microscopy used in the measurement was "QVENTER-200".

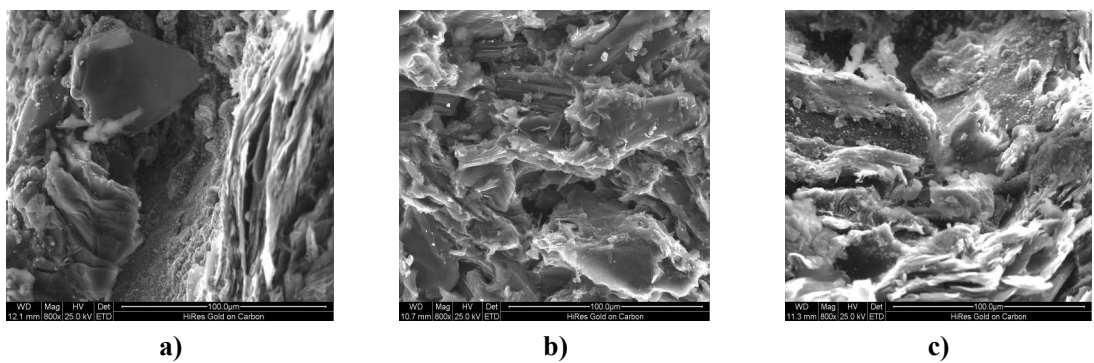

| a) | b) | c) |

Figure 2. Scanning electronic microscopy (SEM) photos of graphite **after synthesis** (x800)

a) Cutting section,   b) Up section,   c) Down section

Let's see the section. It can be seen that the crystallization process, i.e., 3 dimensional arrangements of atoms, at the plane contacting with catalysis was dominant. At the opposite plane the fiber is laid and void is formed between them. The pores of several micron size exists and the single graphite crystal of up to 50 ㎛ size can be seen. And graphite crystals of 10~50 ㎛ size are shown around them.

Next, it can be seen that the particles is crystallized in long shape at the plane contacting with the catalysis, while at the opposite plate the graphitizing was still locally and long arrangement by pressing was seen. It shows that the high pressure and the catalytic metal can facilitate the crystallization of graphite. The shape of graphite is close to the spherical shape in the region of catalyst but the shape at the opposite side is any shape.

The change process of graphite crystal at high pressure is different from the general graphitizing mechanism. At normal pressure, it is difficult to realize the crystallization over 400 ㎚ even though controlling the graphitizing and moreover it needs high temperature over 3000 ℃. The graphite however crystallizes up to several tens of microns. This phenomenon can show that the creation of big inside stress by high pressing at high pressure and high temperature rises the graphitizing, comparing with the previous research result that the graphite crystal contributes to the graphitizing even when the graphite crystal has small thermal expansion coefficient due to the natural behavior of structure at normal pressure.

The bigger stress at the section contacting with the catalyst is generated because the catalytic alloy itself is metal. And because the catalytic materials and graphite have strong reactivity in general, diffusion of metal atoms into the interstitial space between graphite lattices can facilitate such big change, that is, recrystallization. Such recrystallization of graphite up to several tens of microns affects the diamond synthesis.

The catalytic materials for diamond synthesis and behavior of graphite at high pressure and high temperature show that the catalysts and controlling the recrystallization of graphite under super high pressure and high temperature are very important to improve the quality of diamond synthesis.

## 3. Determination of crystal lattice structure and cohesive energy of 3d transition metal carbides by first-principle calculation

We use density functional theory[21] to examine the crystal structure of carbides containing transient metal elements Mn, Fe, Co, and Ni, compare them with experimental data, and then calculate the cohesive energy.

### 3.1. Determination of crystal lattice structure

First of all, we obtained the optimal atomic coordinates using the experimental lattice constants ($a$=0.4526nm, $b$=0.5089nm, $c$=0.6743nm) of cementite $Fe_3C$. The atomic coordinates in cementite structure are shown in Table 1.

Table 1. The atomic coordinates in cementite structure.

| Crystallographic coordinates | | | Atom |
|---|---|---|---|
| $x_1$ | 1/4 | $z_1$ | 4c |
| $-x_1$ | 3/4 | $-z_1$ | |
| $1/2-x_1$ | 3/4 | $1/2+z_1$ | C |
| $1/2+x_1$ | 1/4 | $1/2-z_1$ | |
| $x_2$ | 1/4 | $z_2$ | 4c |
| $-x_2$ | 3/4 | $-z_2$ | |
| $1/2-x_2$ | 3/4 | $1/2+z_2$ | Fe |
| $1/2+x_2$ | 1/4 | $1/2-z_2$ | |
| $x_3$ | $y_3$ | $z_3$ | 8d |
| $-x_3$ | $-y_3$ | $-z_3$ | |
| $1/2+x_3$ | $1/2-y_3$ | $1/2-z_3$ | |
| $1/2-x_3$ | $1/2+y_3$ | $1/2+z_3$ | |
| $-x_3$ | $1/2+y_3$ | $-z_3$ | Fe |
| $x_3$ | $1/2-y_3$ | $z_3$ | |
| $1/2-x_3$ | $-y_3$ | $1/2+z_3$ | |
| $1/2+x_3$ | $y_3$ | $1/2-z_3$ | |

Figure 3 shows the atomic structure of cementite structure. The atomic position parameters to be determined are 7 ($x_1$, $z_1$, $x_2$, $z_2$, $x_3$, $y_3$, $z_3$).

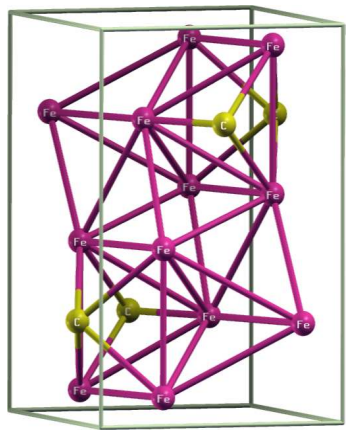

Figure 3. Unit cell of cementite structure.
Fe atoms make the high dense packed structure and carbon atoms are inserted among them, forming the inserting phase.

The atomic coordinate parameters obtained in the simulations are (0.877, 0.440, 0.038, 0.837,

0.176, 0.068, 0.332) in Fe₃C, which are well agreed with the experimental ones (0.890, 0.450, 0.036, 0.850, 0.186, 0.063, 0.328). These values in other carbides Mn₃C, Co₃C, Ni₃C are also well agreed with the experimental ones within the negligible error of $10^{-2}$ magnitude.

Next we performed the crystal lattice optimization. The obtained lattice constants are $a$=0.4514nm, $b$=0.5063nm, $c$=0.6741nm in Fe₃C, which are very close to the experimental ones. The situations are similar in other carbides

We calculated and analyzed the electron density ditribution to make analysis about the nature of chemical bonds in the cementite structure (Figure 4).

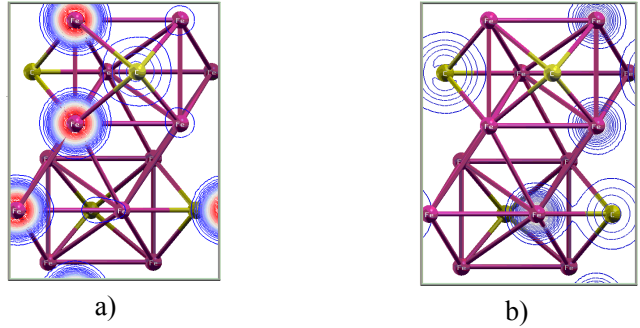

a)        b)

Figure 4. Contour plot of electron density in cementite Fe₃C.
a) Around Fe atom, b) around C atom

As shown in the Figure 4, the electron density is distributed intensively with almost spherical symmetric shape around Fe atom, while small amount of electron density is visible around C atom. Such thing tells us that the electrons transferred from C to Fe atom, Fe atoms are bonded with strong metallic bond and weak ionic bonding between Fe and C atoms exists. The carbon atoms, therefore, can be taken out of crystal.

### 3.2. Determination of binding energies of crystals

After determination of lattice constant parameters, we obtained the cohesive energy by subtracting the total energy of the crystal from the summation of total energies of atoms composing the crystal and dividing it by the number of atoms. Since the structure factor is 4 in the cementite structure, the cohesive energy is as follows,

$$E_{coh} = \frac{1}{16}\left[(12E_M + 4E_C) - E_{M_3C}\right] \quad (1)$$

where $E_M$ and $E_C$ are the total energies of isolated metal atom and carbon atom respectively and $E_{M_3C}$ is one of the crystal composed of 4 M₃C units, that is, unit cells containing 16 atoms. The obtained cohesive energies are shown at Table 2.

Table 2. Cohesive energies of carbides

|          | Cohesive energy (eV/atom) | |
|----------|------|------|
| Carbides | Calc. | Exp. |
| Mn3C | 6.87 | _ |
| Fe3C | 6.53 | 5.05 |
| Co3C | 6.28 | _ |
| Ni3C | 5.75 | _ |

The bond length between carbon and metal atoms in considered $M_3C$-type carbides are 0.204nm and 0.206nm.

Next, we inserted metal atoms into graphite (Figure 5), calculated the total energy and estimated the subtraction between the total energies of isolated metal atom and graphite crystal. In Table 3 the total energies and the subtractions are shown. The total energy of the pure graphite crystal is 22.79431 Ha.

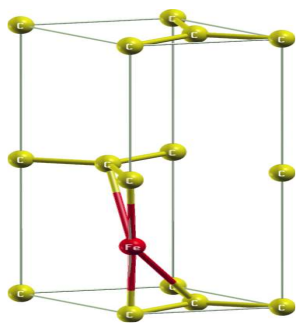

Figure 5. The crystal structure of graphite inserted metal atoms

Table 3. Total energies and the subtraction

| Element | Isolated Atom(Ha) | graphite+metal(Ha) | subtraction(eV) |
|---------|-------------------|---------------------|-----------------|
| Mn | -13.67250 | -36.514519 | 1.30 |
| Fe | -19.37546 | -42.089635 | -2.18 |
| Co | -26.22252 | -48.721306 | -8.04 |
| Ni | -34.59056 | -56.499039 | -24.10 |

1 Ha = 27.2113834 eV = 4.3597425 $10^{-18}$ J

From the Table 3, we can see that the insertion of Mn makes the structure more stable, while other metals make the structure more unstable increasing the atomic number. The atomic radii of Fe, Co and Ni are similar (0.126nm, 0.125nm, 0.124nm).

# 4. Conclusion

At high pressure and high temperature, the recrystallization characteristics of the catalyst metal and the graphite particles appear differently during the diamond synthesis process. When considering the calculation values in the cementite structure of Mn, Fe, Co, and Ni compared with the experimental ones, it is shown that their catalytic behaviors are different as the catalysis for synthesizing diamond. We can fabricate better diamond if the catalysis for synthesizing diamond could be developed considering such behaviors.

From the above consideration, we propose a point of view about the diamond synthesis mechanism by metal catalytic method under extra high pressure.

The graphite that is used in diamond synthesis is not single crystal but the polycrystalline solid consisted of graphite single crystalline grain with differenct sizes joined as amorphouse layer. When increasing the temperature under the extra high pressure, the catalytic metal atoms close to melting point diffuse through layers of the graphite single crystalline grain, forms the seed of diamond crystal and thus graphite becomes the polycrystalline solide consisted of both diamond seeds and graphite single crystalline grain not transformed into diamond. When such polycrystalline solid is resolved into metal melts, it does not form the true solution of carbon atom but the amorphouse layers are firstly resolved as atomic state and then it is resolved slowly forming the colloid system by diffusion of the diamond seeds and graphite single crystalline grains. After all, graphite-catalytic metal melt consists of the true solution of carbon and diamond grain and graphite colloid grains drifting in it. Then the catalytic metal ions diffuse into the graphite lattice layers in the graphite colloid grains and transformed into the seeds of dimaond crystal by facilitating the martensite transformation. The diamond crystal is grown by attachement of the carbon atom in melt or carbon micro cluster.

Here the catalytic metal should be face centered cubic crystal structure of the lattice constant similar with cubic diamond, the bond length between the catalytic metal ion and carbon atom should be smaller than 0.207 nm and the strength of binding should be high and be dissociated easily. Above mentioned theorectical consideration is the theoretical prove on our such view of point.

## Acknowledgments

It is a pleasure to thank Kwang Il Jon, Jae Sik Jang and Nam Hyok Kim for useful discussions. This work was supported by the National Program on Key Science Research of Democratic People's Republic of Korea (Grant No. 18-1-5).